\documentclass[11pt,twoside]{article}
\usepackage{asp2010}

\resetcounters

\begin{document}

\title{On the CQE Phase Signature Feature in High-Resolution Spectro-Interferometry of Be Stars}
\author{D. M. Faes,$^{1,4}$ A. C. Carciofi,$^1$, Th. Rivinius$^2$, S.~\v{S}tefl$^2$, D. Baade$^3$, A. Domiciano de Souza$^4$}
\affil{$^1$Instituto de Astronomia, Geof\'isica e Ci\^encias Atmosf\'ericas, Universidade de S\~ao Paulo, Rua do Mat\~ao 1226, Cidade Universit\'aria, 05508-900, S\~ao Paulo, SP, Brazil, moser@usp.br, carciofi@usp.br}
\affil{$^2$European Organisation for Astronomical Research in the Southern Hemisphere, Casilla 19001, Santiago 19, Chile}
\affil{$^3$European Organisation for Astronomical Research in the Southern Hemisphere, Karl-Schwarzschild-Str. 2, 85748 Garching bei M\"unchen, Germany}
\affil{$^4$Lab. J.-L. Lagrange, UMR 7293 - Observatoire de la C\^{o}te d'Azur (OCA), Univ. de Nice-Sophia
Antipolis (UNS), CNRS, Valrose,  06108 Nice, France}

\begin{abstract}
We analyse a feature that is detectable at the differential phases of high-resolution spectro-interferometry of Be stars.
The origin of this feature (dubbed CQE-PS, Central Quasi Emission Phase Signature) lies in the differential absorption of photospheric radiation by the circumstellar disk, which is spectroscopically detected as a CQE line profile in shell stars. This phenomenon has great diagnostic potential for Be star disks, revealing properties of these disks on the scale of a few stellar radii. 
\end{abstract}

\section{Introduction} \label{intro}

Classical Be stars are well known to be characterized by having gaseous circumstellar disks that are fed from mass lost from their rapidly rotating central stars \citep{por03}. In the past two decades, optical and infrared interferometry played a key role in the study of these astrophysical systems. For instance, the pioneering measurements of \citet{qui97} provided strong evidence that the circumstellar matter around Be stars is distributed in a geometrically thin structure (i.e., a disk). More recently, detailed measurements allowed to demonstrate that Be disks rotate in a Keplerian fashion \citep{mei07,kra11,whe12}.  

The strong developments in the observations were accompanied by significant progress in our theoretical understanding of these systems. Recently, a consensus is emerging that the viscous decretion disk model (VDDM), first suggested by \cite{lee91} and further developed by Bjorkman (1997), and Okazaki (2001), among others, is the best candidate to explain the observed properties of Be disks \citep{car11,mcg11}.
This model uses the angular momentum transport by turbulent viscosity to lift material that was ejected by the star into higher orbits, thereby causing the disk to grow in size. A theoretical result of this model is that material is in Keplerian rotation throughout the disk, in agreement with the recent interferometric results.



Before the spectro-interferometric era, one of the first strong indications that Be disks are Keplerian came from high-resolution spectroscopy ($R>30000$) of shell stars. \citet{han95} and \citet{riv99} studied the so-called \textit{Central Quasi-Emission} (hereafter, CQE) peaks, where the disk, under certain circumstances, causes a cusp in the deepest region of the line profile of shell stars. The existence of the CQE implies slow radial motions of the gas, which means the disk is supported by rotation.

In this work, we show that the same mechanism that produces  the CQE observed in shell star line profiles can cause important changes in the intensity map of the Be star plus disk system, with observable effects on the interferometric quantities. We analyze and quantify the diagnostic potential of this effect, which we dub \textit{CQE Phase Signature} (hereafter, CQE-PS). Interestingly, similarly to the CQE detection, the CQE-PS can only be studied by high-resolution observations ($R>12000$), such as can be obtained by AMBER/VLTI in the near infrared.



\section{Model Description} \label{refcase}

We will see in Sect.~\ref{prfeat} that the mechanism behind the CQE-PS is the differential absorption of photospheric radiation by the circumstellar disk. This mechanism can be directly traced by the interferometric differential phase, and must be separated from other mechanisms that affect this quantity.

To study, characterize and illustrate the CQE-PS, we adopt a realistic model for the Be + disk system. 
For the central star we adopt a rotationally deformed and gravity darkened star whose parameters are typical of a B1Ve star (Table~\ref{tab:bemod}). In this study, the only free parameter of the star is its rotation rate ($\Omega/\Omega_{\rm crit}$).
For the circumstellar disk, we adopt the VDDM. 
For the case of isothermal viscous diffusion in the steady-state regime, a state reached after a sufficiently long and stable decretion period (Haubois, these proceedings), the disk surface density assumes a particularly simple form $\Sigma(r) \propto r^{-2}$ \citep{bjo05}.
The interferometric quantities depend on the spatial resolution with which the object is seen. To explore different configurations we define the quantity $\nu_{\rm obs}$, the ratio between the distance to the star (i.e., its angular size on sky) and the baseline length of the interferometer. The units used are m\,pc$^{-1}$.

\begin{table}[!ht]
\begin{center}
\caption{Reference Be model parameters.}
{\small
\begin{tabular}{ccc}
\noalign{\smallskip}
\hline 
\hline 
\noalign{\smallskip}
Parameter & Symbol & Ref. Case \\
\noalign{\smallskip}
\hline 
\noalign{\smallskip}
Spectral type & - & B1V \\
Polar radius & $R_{\rm pole}$ & 4.9 $R_{\odot}$\\
Pole temperature & $T_{\rm pole}$ &  27440 K  \\
Luminosity & $L_{\rm *}$ & 10160 $L_{\odot}$ \\
Critical velocity & $v_{\rm crit}$ & 534.4 km/s \\
Rotation rate & $\Omega/\Omega_{\rm crit}$ & 0.8 \\
Oblateness & $R_{\rm eq}/R_{\rm pole}$ & 1.14 \\
Gravity darkening & $T_{\rm pole}/T_{\rm eq}$ & 1.16  \\ 
\noalign{\smallskip}
\hline 
\noalign{\smallskip}
Disk radius & R$_{\rm disk}$ & 10 R$_{\rm *}$  \\
Disk density & $n$ & 10$^{13}$g cm$^{-3}$ \\
\noalign{\smallskip}
\hline
\noalign{\smallskip}
Incl.angle & $i$ & 45$^\circ$  \\
Spectral res. & R & 12000  \\
Baseline/distance & $\nu_{\rm obs}$ & 0.5 m pc$^{-1}$ \\
Position angle & PA & 0$^\circ$ \\
\noalign{\smallskip}
\hline
\end{tabular} }
\label{tab:bemod}
\end{center}
\end{table}

\subsection{Interferometric Phases in the Marginally-Resolved Case} \label{canon}

We see below that the CQE-PS at interferometric differential phases (hereafter DP) happens under quite specific circumstances, and its diagnostic potential is more relevant for some parameters than for others. To quantify this, we first study a reference model for which the CQE-PS is not present, i.e., a model for which the canonical signatures for interferometric quantities apply. The adopted model parameters for the reference case are listed in the third column of Table~\ref{tab:bemod}.

As pointed out by \citet{dom04}, an image $\imath(\textbf{r},\lambda)$ of an astronomical target is defined as the convolution of the intensity map of the object, $I(\textbf{r},\lambda)$, with the Point Spread Function (PSF) due to the instrument and the terrestrial atmosphere, $A(\textbf{r},\lambda)$, i.e., 
$\imath(\textbf{r},\lambda)=I(\textbf{r},\lambda)*A(\textbf{r},\lambda), $
where $\textbf{r}$ is the position on the sky vector.

When the data is obtained by an interferometer, the vector $\textbf{u}$ represents the spatial frequency associated with $\textbf{r}$. In long baseline interferometry, $\textbf{u}$ is the vector $\textbf{B}_{\rm proj}/\lambda$. At high spectral resolution 
the interferometric phase signal $\phi$ can be obtained as an expansion of $\tilde{I}(\textbf{u},\lambda)$, the Fourier transform of the brightness distribution.

If the target is not resolved $\textbf{u}\cdot\textbf{r} \ll 1$, i.e., very small values of $\nu_{\rm obs}$, the differential phase can be written as \citep[Eq. 4]{dom04}
\begin{equation}
\phi(\textbf{r},\lambda, \lambda_r)=-2\pi\textbf{u}\cdot[\epsilon(\lambda)-\epsilon(\lambda_r)] \,;
\label{eq:photc}
\end{equation}
the $\epsilon(\lambda)$ and $\epsilon(\lambda_r)$ vectors are called \textit{photocenters} respectively for $\lambda$ and $\lambda_r$, where $\lambda_r$ is a reference channel observed simultaneously. The usual procedure is to take $\lambda_r$ on the adjacent continuum of the spectral line, with $\epsilon(\lambda_r)=0$. 

\begin{figure}[!t]
\centering
\includegraphics[width=.9\linewidth]{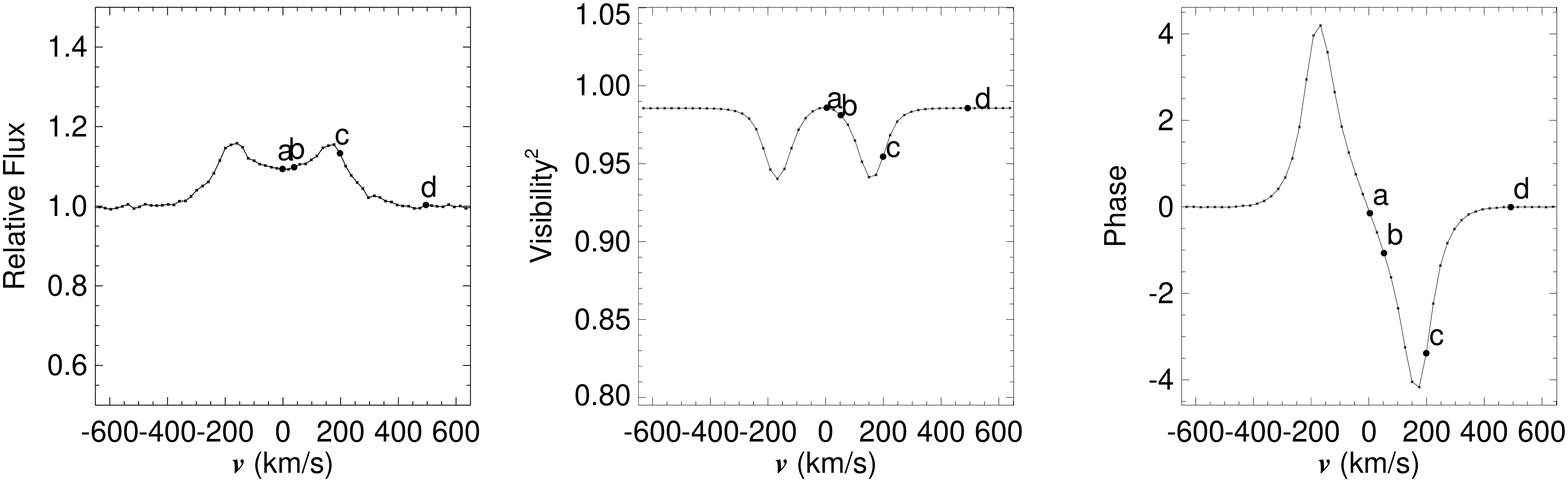}\\
\includegraphics[width=.75\linewidth]{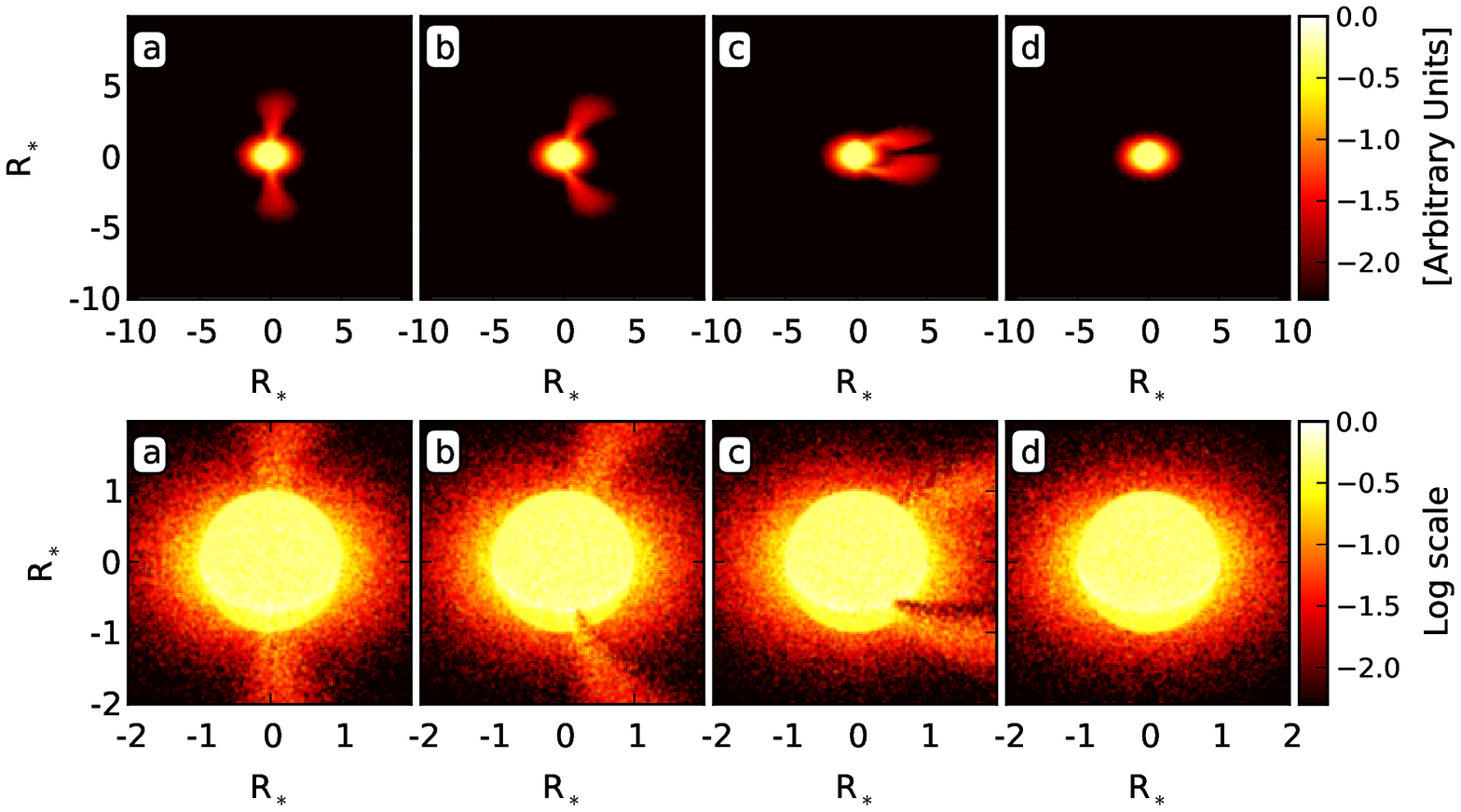}
\caption{\emph{Top 3 panels}: synthetic line flux, interferometric visibilities and phases across the Br-$\gamma$ line for the reference model. Calculations were made for $\nu_{\rm obs}=0.5\,\rm m\,pc^{-1}$, and baseline orientation parallel to the disk equator.
\emph{Bottom panels:} The Be model images for the reference model associated with its high-resolution interferometric measurements. 
The disk inclination angle is $i=45^\circ$.}
 \label{fig:classbe}
\end{figure}

For the case of rotating disks, it is usually assumed that the DP will have a simple S-shaped profile \citep{ste96}.
This can be easily understood from Fig.~\ref{fig:classbe}, which shows the Br-$\gamma$ line profile, the corresponding interferometric quantities and model images at different wavelengths for the reference model. Shown are results for $i=45^\circ$. At continuum wavelengths (d, in Fig.~\ref{fig:classbe}) the images show the star surrounded by a  centrosymmetric continuum emission. We define the DP to be zero at those wavelengths.
As we move from the continuum towards line center (d to a) the line flux initially increases as a result of progressively larger emission lobes and then decreases as the emission area decreases towards line center (a). This creates the familiar W-shaped pattern in the visibilities and the S-shape pattern for the DPs.





\subsection{Interferometric Phases in the Resolved Case} \label{nr_interf}

The association between the phase signal and the image's photocenter position is only possible if the marginally-resolved condition is satisfied, since it permits the truncation of the Fourier expansion \citep{jan01}. This is particularly important for Be stars since their on-sky size is wavelength dependent, mainly across emission lines where the size of the emitting regions varies with wavelength.

\begin{figure}[!t]
\centering
\includegraphics[width=.35\linewidth]{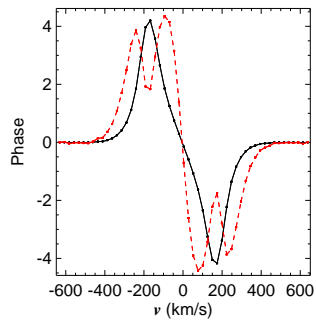}
\caption{Differential phase signal for the Be reference model. The reference model is the black (full) line, observed with $\nu_{\rm obs}= 0.5\,\rm m\,pc^{-1}$ (astrometric regime). In the red (dashed) line, $\nu_{\rm obs}=2.5\,\rm m\,pc^{-1}$ (complex phase behaviour).}
 \label{fig:unres}
\end{figure}

\begin{figure}[!t]
\centering
\includegraphics[width=.6\linewidth]{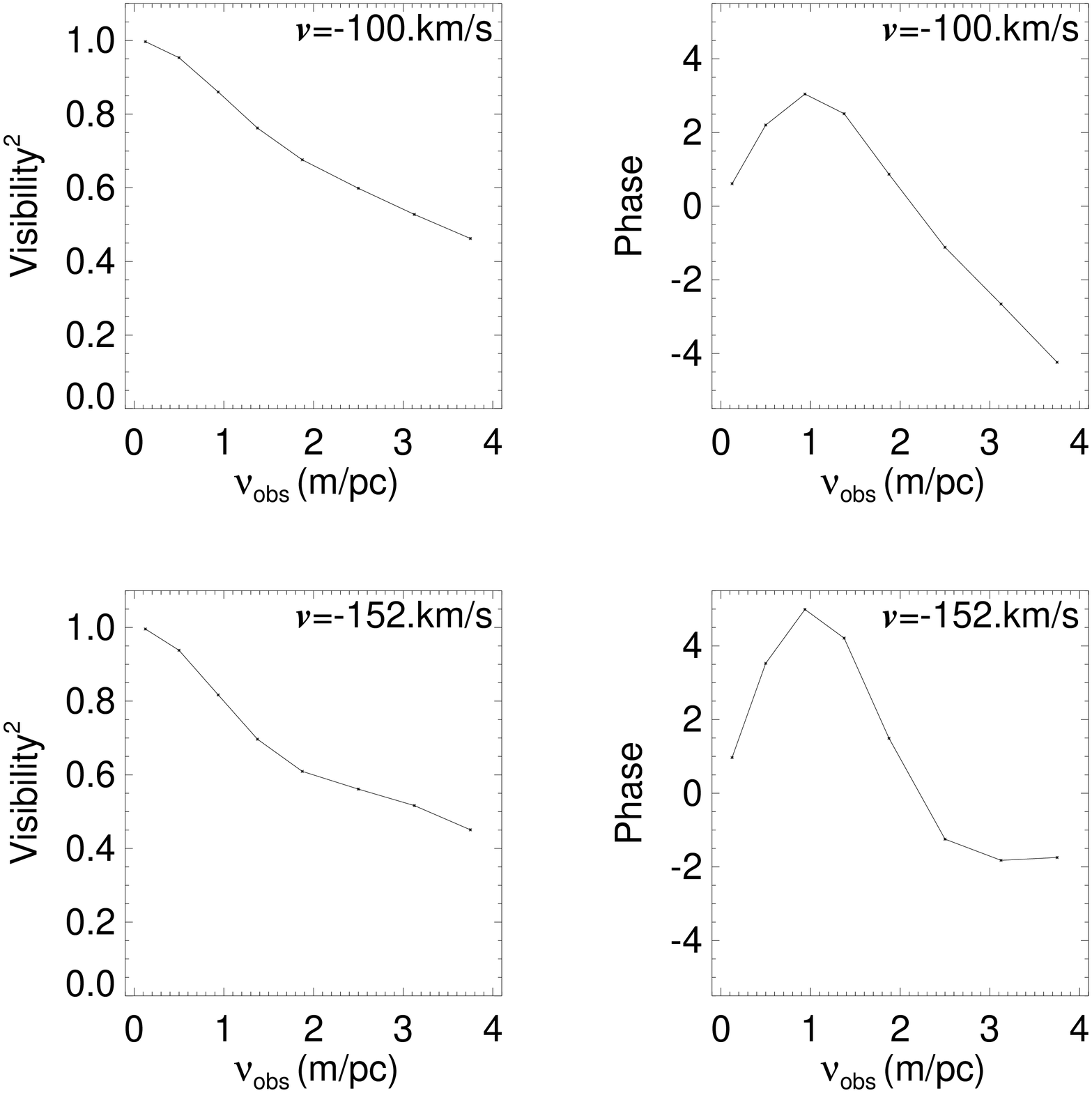}
\caption{The squared visibility and differential phases as function of interferometric baselines for the Be star reference model at radial velocities of $v=-100\rm\,km\,s^{-1}$ and $v=-152\rm\,km\,s^{-1}$.}
 \label{fig:interf}
\end{figure}

Outside the astrometric regime (i.e., $\textbf{u}\cdot\textbf{r} \approx 1$ or large values of $\nu_{\rm obs}$), Eq.~(\ref{eq:photc}) is no longer valid. A complex phase behavior is then expected since higher order terms of the Fourier transform of the image are relevant, and it does not correspond anymore directly to the photocenter.
Fig.~\ref{fig:unres} compares the DP signal of our reference model in the astrometric regime and outside it.
The reason the curves are so different is explained in Fig.~\ref{fig:interf} that plot the interferometric quantities for two wavelengths accross Br-$\gamma$ as a function of baseline length. The complex phase behavior, even with change of sign, occurs even before the fully resolved regime, i.e., when the squared visibility goes to zero.

Based on our simulations, the astrometric regime is approximately valid up to squared visibilities of $\approx0.8$, which corresponds to $\nu_{\rm obs} \lesssim 1.5\rm\,m\,pc^{-1}$ for a typical Be star. The exact value of these quantities is dependent of the target brightness distribution.
One example of the complex behavior outside the astrometric regime is reported by \citet{kra11}, who models AMBER/VLTI data for the Be star $\beta$~CMi. In this case, a phase reversal is registered at the DP signal of the longest baseline available for this nearby target (squared visibility of $\approx0.7$, $\nu_{\rm obs}\approx2.2$).

Outside the astrometric regime a phase reversal is an example of the phenomenology that emerges entirely from long baseline interferometry, where the phase signal no longer corresponds to displacements of the photocenter. In the text that follows all interferometric computations are in the marginally resolved regime to adequately separate the CQE-PS phenomenon from the just described effects.

\section{The CQE Phase Signature} \label{prfeat}
For the reference model described in Sect.~\ref{canon}, line emission from the rotating disk is the most important factor controlling the interferometric quantities. However, in some circumstances line absorption of photospheric light by the disk can also have a strong impact on these observables.

The spectroscopic phenomenon of the CQE was first studied by \citet{han95} for the so-called shell absorption lines in nearly edge-on Be shell stars. These consist of narrow absorption lines well below the stellar continuum that show a central weak cusp, known as Central Quasi-Emission peaks (CQEs). As pointed out by \citet{riv99}, these peculiar profiles result from the minimum at zero radial velocity in the absorption of the photospheric flux by the rotating circumstellar gas. 

CQEs are, in spite of their name, not related to any emission process but are a pure absorption phenomenon. So, when observing the star plus disk system the latter acts as an absorber of the stellar photospheric light, as seen in the CQE spectroscopy profile. 

\begin{figure}[!ht]
\centering
\includegraphics[width=.9\linewidth]{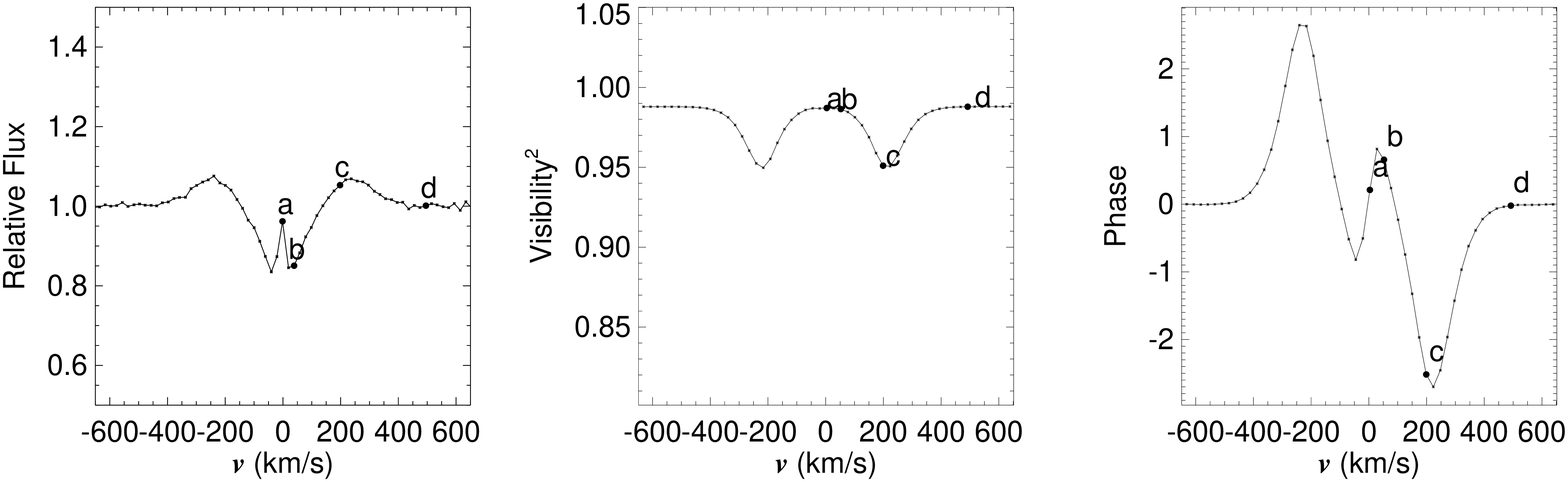}\\
\includegraphics[width=.75\linewidth]{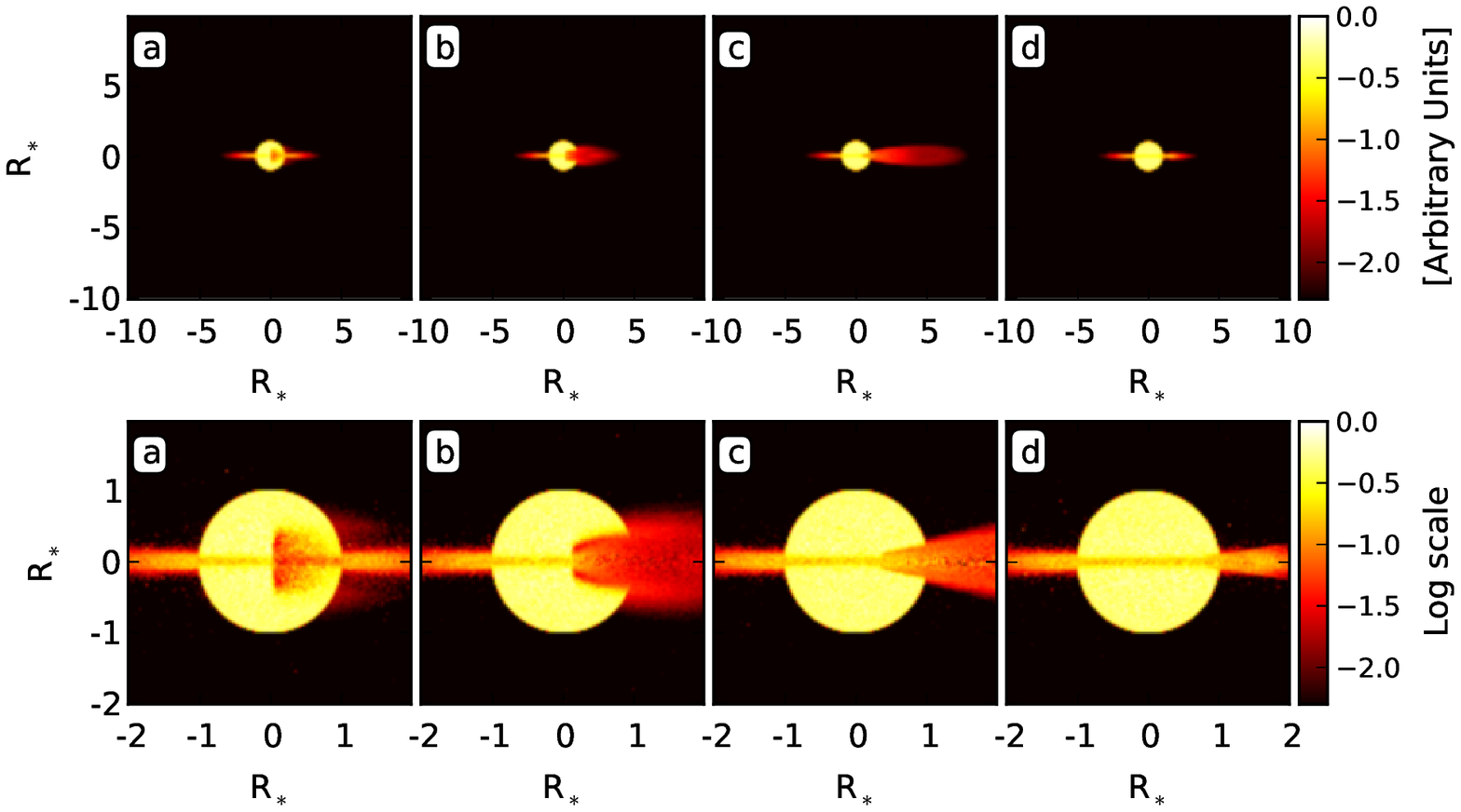}
\caption{\emph{Top 3 panels}: synthetic line flux, interferometric visibilities and phases across the Br-$\gamma$ line for the reference model. Calculations were made for $\nu_{\rm obs}=0.5\rm\,m\,pc^{-1}$, and baseline orientation parallel to the disk equator.
\emph{Bottom panels:} The Be model images for the reference model associated with its high-resolution interferometric measurements. 
The disk inclination angle is $i=90^\circ$.}
 \label{fig:phot}
\end{figure}

The differential absorption across the spectral line that causes the CQE also has strong effects on interferometric differential phases, as shown in Fig.~\ref{fig:phot}, where we plot results for the reference case seen at $i=90^\circ$ (edge-on). The projected absorption in the stellar disk generates a considerable decrease of the stellar flux, which in turn will affect the photocenter of the system, with a corresponding signal in the DP.


Consider the system seen at continuum wavelengths (Fig.~\ref{fig:phot}d). The dark lane across the stellar disk is caused by continuum (bound-bound and bound-free) absorption. As we consider images at different spectral channels across the Br-$\gamma$ line, we also see line (bound-bound) absorption of photospheric light and emission by the disk (Figs.~\ref{fig:phot}a to \ref{fig:phot}c). This line absorption has a differential aspect because it depends on the line of sight velocity of the absorbing material.

At high velocities, only a small area near the edge of the disk will block the starlight (Fig.~\ref{fig:phot}c). For lower velocities, an increasing disk portion will block the starlight, creating a maximum absorption around 50\,km\,s$^{-1}$ (Fig.~\ref{fig:phot}b). Finally, as one considers wavelengths closer to the line center, the disk absorption decreases due the fact that there is less disk material with the appropriates line-of-sight velocity. 

\begin{figure}
\centering
\includegraphics[width=.5\linewidth]{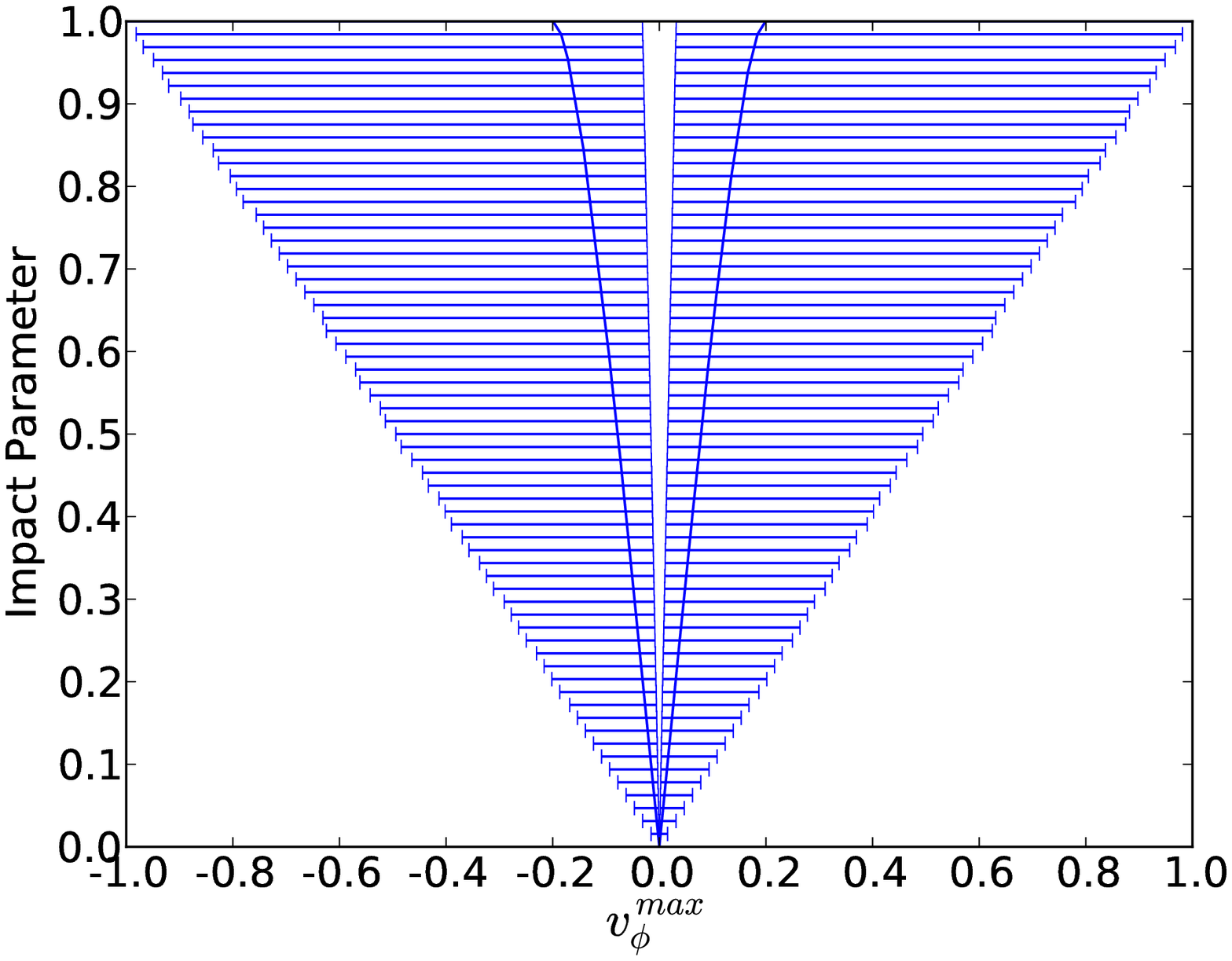}
\caption{Projected velocity range of a rotating disk  as a function of impact parameter. The solid line shows the average geometric velocity component for each impact parameter for a Keplerian disk of $R_{\rm disk} = 10\,R_{\star}$. Disk inclination angle is $i=90^\circ$.}
 \label{fig:lambda_vs_p}
\end{figure}

The CQE effects are concentrated very near the central wavelength, since there the opacities along the line-of-sight are largest. Fig.~\ref{fig:lambda_vs_p} shows  the  projected velocity interval of the disk  as a function of the stellar impact parameter. The low-velocity components are present for all impact parameters, and that is why the CQE-PS affects more the DP close to line center.

Contrary to disk emission that shifts the photocenter position towards the direction of the emitting lobe, line absorption of photospheric light shifts it on the opposite direction. So, when evaluating the differential phases, we are measuring the competition between the photocenter shift due to disk emission vs. the shift due to disk absorption. If absorption is strong enough, or emission is weak (for instance, as expected for a small disk density) one may observe a reversal in the DP. 

\section{Conclusions} \label{concl}
The new interferometric observations of Be stars at high spectral resolution lead to new measurements of differential phases. Deviations from the canonical appearance may occur in two different circumstances: (i) by purely interferometric properties at observations outside the marginally resolved regime, where the differential phases no longer correspond to the photocenter of the system, and (ii) by physical process accessible at high-resolution affecting the interferometric signal. In particular, in this work we discuss the role of the disk differential absorption as an important factor in shaping the differential phases of Be stars.




The solution of the radiative transfer problem of stellar photospheric flux in interaction with a typical rotating Be disk can cause a measurable photocenter displacement. Since this is an effect of differential absorption by the disk, it will depend on various system parameters and can provide valuable information about its configuration.
The diagnostic potential of the CQE-PS in determining important disk parameters, such as density and velocity field, will studied in a future publication (Faes et al., in prep.).


\bibliography{talk_faes_bib}

\question{Millour} Your ``CQE effect'' is a physical effect which should be seen at all baselines lengths, including spectro-astrometric measurements on Be stars. On the other hand, a resolution effect (spatially speaking) would produce a ``phase reversal'' which depends on the baseline length, i.e. it would be seen only on longest baselines and not on spectro-astrometry data. My experience would make me point to that hypothesis. What is yours?

\answer{Faes} Indeed long baselines lead to complex phase signals - including a reversal. We were aware of this in our simulations. The phase reversal presented is assigned to photocenter changes from a physical cause, namely stellar light absorption, that occurs for specific circumstellar disk configurations.

\end{document}